\documentclass[12pt, preprint]{aastex}

\shortauthors{Ji et al.
}
\shorttitle{}
\begin{document}

\title{Early Abnormal Temperature Structure of X-ray Looptop Source of Solar Flares}
\author{Jinhua Shen\altaffilmark{1,2}, Tuanhui Zhou\altaffilmark{2}, Haisheng Ji\altaffilmark{2,3}, Na Wang\altaffilmark{1}, Wenda Cao\altaffilmark{3}, Haimin Wang\altaffilmark{3}}

\altaffiltext{1}{Urumqi Observatory, National Astronomical Observatories, Urumqi, 830011, China}
\altaffiltext{2}{Purple Mountain Observatory, 2 West Beijing Road, Nanjing,
210008, China}
\altaffiltext{3}{Center for Solar Terrestrial Research, New Jersey Institute of
Technology, University Heights, Newark, NJ, 07102, USA}

\begin{abstract}
This Letter is to investigate the physics of a newly discovered
phenomenon --- contracting flare loops in the early phase of solar
flares. In classical flare models, which were constructed based on
the phenomenon of expansion of flare loops, an energy releasing site
is put above flare loops. These models can predict that there is a
vertical temperature gradient in the top of flare loops due to heat
conduction and cooling effects. Therefore, the centroid of an X-ray
looptop source at higher energy bands will be higher in altitude,
for which we can define as normal temperature distribution. With
observations made by {\it RHESSI}, we analyzed 10 M- or X-class
flares (9 limb flares). For all these flares, the movement of
looptop sources shows an obvious U-shaped trajectory, which we take
as the signature of contraction-to-expansion of flare loops. We find
that, for all these flares, normal temperature distribution does
exist, but only along the path of expansion. The temperature
distribution along the path of contraction is abnormal, showing no
spatial order at all. The result suggests that magnetic reconnection
processes in the contraction and expansion phases of these solar
flares are different.

\end{abstract}


\section{Introduction}
It is widely accepted that solar flare energy comes from sudden
release of free magnetic energy via magnetic reconnection. In
commonly adopted flare models, the ever-ascending Y-type
reconnection point in the solar corona results in expanding flare
loops and separation motion of flare foot points (FPs) (Kopp \&
Pneuman 1976). The expansion of flare ribbons and flare loops is an
important signature of progressive magnetic reconnection in the
corona. However, the contraction of flare loops in the early phase
of flares may suggest a different reconnection scenario.

The signature for the contraction of flare loops has been reported
by several authors (Sui et al. 2003, 2004; Li \& Gan 2005; Liu et
al. 2004; Veronig et al. 2006; Ji et al. 2004b, 2006, 2007, 2008).
Contraction motion includes two aspects: the converging motion of
FPs and the correlated downward motion of a LT source, such as the
M1.1 flare of 2004 November 1 (Ji et al. 2006). The contraction
picture is different from the shrinkage of flare loops with rooted
FPs being fixed or still in expansion. From our experience, the
similar event like the M1.1 flare is rare, since HXR FPs are usually
missing during the initial phase of a flare, such as the X3.9 flare
of 2003 November 3 (Veronig et al. 2006). To investigate the
converging motion of FPs, the most preferable wavelength is
H$_\alpha$ blue wing, at which H$_\alpha$ emission is believed to be
caused by nonthermal electrons (Canfield et al. 1984).

The classical flare model puts the energy releasing site above flare
loops, from which we can expect that a higher temperature source
will be located above a lower temperature source due to heat
conduction and cooling effects. In this Letter, we will name this
distribution as normal temperature distribution (NTD). For the X10
flare of 2003 October 29, Ji et al. (2008) reported that a HXR
sigmoid structure contracts during the impulsive phase of the flare.
The contraction is the result of reconnection between two
highly-sheared flux ropes. For magnetic reconnection between flux
ropes, the existence of NTD is not required, since the energy
releases inside flux ropes. Thus, temperature distribution along the
altitude is an important factor for testing the reconnection
scenarios.

Based on above thinking, we re-analyzed the M2.1 flare of 2002
September 9. The flare is a well-observed sample showing the early
converging and subsequent separation motion of the flare kernels. We
found that the motion of the HXR LT source is well correlated with
that of the flare kernels. Notably, the NTD occurs only during the
expansion period of flare loops. During the contraction period,
higher and lower temperature structures are mixed together. To find
supporting evidences for this abnormal temperature distribution
during the contraction period, we surveyed nearly all M-class and
X-class limb flares from 2002 to 2005, which were well-observed by
{\it RHESSI} (Lin et al. 2002), we found at least 9 events showing
the phenomenon.

In \S 2 we present the result of the flare of 2002 September 9, in
\S 3 we give the other 9 supporting events. Discussions and
conclusions are briefly given in \S 4.

\section{The flare of 2002 September 9}

The M2.1 flare occurred on 2002 September 9, starting at 17:40. The
flare was well observed at Big Bear Solar Observatory at
H$_\alpha$-1.3{\AA} with very high time resolution (40 ms). By
analyzing the flare with detailed spatial, spectral and temporal
information, Ji et al. (2004a) identified nonthermal and thermal
(multiple temperature) HXR spikes. They further reported that the
distance between the two H$_\alpha$ conjugate kernels decreases
during the impulsive phase of the flare. Only after the impulsive
phase, the distance has a steady increase showing the usual
separation motion (Ji et al. 2004b). They also mentioned that the
height of the HXR looptop source decreases during the impulsive
phase. The flare has two pair of conjugate kernels in H$_\alpha$
blue wing (a1-a2, and b1-b2 in Figure 1). From Figure 1a, we can
clearly see two EUV loop systems connecting a1 to a2 and b1 to b2
respectively. Judged from temporal and spatial relationship between
HXR emission and H$_\alpha$ emission, HXR emissions of this event
come dominantly from the flare loops connecting a1 to a2.
Furthermore, there is an over 1 minute delay for kernels b1 and b2
and the magnetic reconnection producing b1 and b2 was induced by the
magnetic reconnection producing a1 and a2 via loop interactions
(Huang \& Ji 2005).

To study the moving behavior of the HXR LT source, we constructed
{\it RHESSI} CLEAN HXR maps in the energy ranges of 8-10, 10-13, and
13-16 keV using 16-s time bin. HXR emissions in these energy ranges
originate from the top of the EIT 195 \AA\ flare loops connecting a1
and a2 (Figure 1). The CLEAN maps were made with natural weighting
using grids 3-8, in which the maps have an FWHM spatial resolution
of $\sim$ 10 arcsec. Note that the emission centroids can be
determined with an accuracy of $<$ 1 arcsec, depending on the count
statistics (Hurford et al. 2002). We measured the LT source's
centroid positions within the contour level of 50\% using a tool
provided by {\it RHESSI} software. The temporal series of centroid
positions of the LT source in the energy range of 10-13 keV is shown
in Figure 1b, in which red arrows show downward moving direction and
blue ones show upward motion. We measured the distance from the
centroid of the LT source to the base of the flare loops. Since the
flare loops are clearly seen, the distance can be reasonably
regarded as the projected height. The time profiles for the `height'
measured at three different energy ranges are plotted in Figure 2c.
For comparison, the time profile of HXR emission in the energy range
25-50 keV and the distance between the two H$_\alpha$ kernels are
given in Figure 2a-b. We can find that the motion of the HXR LT
source is well correlated with the relative motion between the two
H$_\alpha$ kernels. This confirms the picture of early contraction
and subsequent expansion of flare loops, as summarized by Ji et al.
(2007).

During the expansion period, the temperature structure shows an NTD
(Fig. 2c). This is in agreement with what the standard flare model
predicts with an energy releasing site above flare loops. However,
during the contraction period, the temperature structure of the LT
source is rather complex or abnormal. The LT sources in the three
energy bands are almost mixed with one another. The abnormal
temperature distribution in the contraction period obviously
suggests a complex magnetic reconnection process.

\section{Supporting observations}

To find supporting evidences, we surveyed 70 limb flares well
observed by {\it RHESSI} from 2002 to 2005, most of them are M-class
or X-class. For each event, we constructed {\it RHESSI} Clean maps
in the energy ranges of 8-10, 10-13, and 13-16 keV with a time bin
of 12 seconds. HXR emissions at these energy ranges usually come
from flares' LT. We selected those events with a single LT source
and the LT source exhibits an obvious U-shaped trajectory (only one
turning point), which is assumed to be the signature of early
downward motion and subsequent upward motion. An additional
requirement is that the downward motion should last more than 1
minutes. According to this criteria, 12 events could be selected for
the study including the three homologous flares investigated by Sui
et al. (2004). In this paper, we will not include the three
homologous flares (see Table 1 for the nine flares).

Even for limb events, it is hard to measure the actual height of a
HXR LT source with respect to solar photospheric surface. Actually,
as Sui et al. (2003, 2004) noted, the motion of LT sources of a limb
event is usually not in radial direction. A reference point is
needed to estimate the height of a HXR LT source. For the flare of
2002 September 9, the reference point is chosen as the center of the
line connecting the two FPs of the flare loop. For each of the 9
events, a turning point around the base of the U-shaped path is
taken as the reference point, from which the `height' of a LT source
is estimated. The trajectories of LT sources of the nine events are
plotted in the top panels of Figures 3-4, in which nine ``+'' signs
show the positions of reference points for the events. We can see
that before and after the turning point the trajectories of the LT
sources can be roughly fitted by two straight lines.

The time profiles for the `height' of the LT sources at different
energy ranges are plotted in the middle panels of Figures 3-4. For
comparison, the light curves of corresponding event in different
energy ranges are plotted in the lower panels. For all nine events,
we can find that well-established NTD does occur but only in the
expansion phases. During the contraction periods, high energy and
low energy sources are spatially mixed. It is worth mentioning that
the height curve in the first column of Figure 4 for the flare of
2003 November 3 has been reported by Veronig et al. (2006). Figures
3-4 conform the results that we obtained for the 2002 September 9
event. Turning points occur during the rising phases of the flares.
To some events turning points coincide with peak times of the light
curves.

\section{Discussion and Conclusions}

In this Letter, we report the finding of the early abnormal
temperature distribution of flares' X-ray LT sources during
contraction phase of ten solar flares. In a 2D framework for flare
models, an energy releasing site is assumed to be above flare loops.
Following the scenario, we can expect that a LT source at higher
energies will be higher in altitude due to heat conduction and
cooling effects. The results from the ten flares show that this
expectation can be met, but only during the expansion phase of solar
flares. During the contraction period, however, high energy and low
energy LT sources are mixed, showing a kind of abnormal temperature
distribution.

The physics for the contraction of flare loops is still not
well-understood. From above ten events, we have seen that the
turning points in the trajectories of the LT sources occur near
flare peak times. Therefore, the contraction is related to magnetic
reconnection in the impulsive phase of solar flares. Ji et al.
(2008) reported that, during the contraction period of the X10 flare
of 2003 October 29, HXR emissions at all energies share the similar
sigmoidal configuration. The contraction corresponds to the
shrinkage of the HXR sigmoid, which is the result of magnetic
reconnection between highly-sheared flux ropes. We may propose that
the abnormal temperature distribution is associated with magnetic
reconnection between highly-sheared flux ropes. On the other hand,
the downward and upward phases of the LT motion may reflect two
regimes of reconnection: bursting and no-bursting (Karlick\'{y}
2008: private communication). The abnormal temperature structure can
be explained by a presence of small plasmoids in the current sheet
just above contracting arcade. The plasmoids are formed in the
bursting regime of the reconnection and move also downwards where
they interact with the arcade. This interaction represents
additional reconnection that changes the normal temperature
structure (B\'{a}rta et al. 2008; Koloma\'{n}ski \& Karlick\'{y}
2007). In the phase of the upward LT motion the reconnection take
place without these plasmoids (no-bursting regime of the
reconnection). Therefore the temperature is in agreement with the
prediction made by standard flare models.

\acknowledgements {This work is supported by CNSF 10473024, CNSF
10333030, 973 project with No. 2006CB806302 and US NASA under grants
NNX 07 AH 78G and NNX 08 AQ 90G}

\clearpage

\begin{table}[h]
\begin{center}
\caption{Events List}
\bigskip
\begin{tabular}{cccccccc}
\tableline
\tableline  Date & Class   &  Starting  & Peak Time    & Turning &  Descending & Expanding  \\
                 &         &  Time      &  Time$^*$        & Time & Speed (km$s^{-1}$) & Speed (km$s^{-1}$) \\
\tableline
\tableline  2002-09-09 & M2.3  & 17:40  & 17:49 & 17:48 & 18.3  & 6.8   &   \\
\tableline  2002-02-20 & M4.3  & 09:46  & 09:58 & 09:54 & 19.2  & 6.8   &   \\
\tableline  2002-04-04 & M6.1  & 15:24  & 15:30 & 15:30 & 33.5  & 19.2  &   \\
\tableline  2002-07-06 & M1.8  & 03:24  & 03:33 & 03:32 &  8.2  & 8.3  &   \\
\tableline  2002-07-20 & X3.5  & 20:52  & 21:08 & 21:08 & 32.3  & 8.1   &   \\
\tableline  2002-07-23 & X4.8  & 00:17  & 00:28 & 00:23 & 12.3  & 15.6  &   \\
\tableline  2003-11-03a & X3.9 & 09:43  & 09:49 & 09:49 & 14.7  & 15.7  &   \\
\tableline  2003-11-03b & X2.9 & 01:01  & 01:17 & 01:16 & 14.8  & 6.9  &   \\
\tableline  2004-08-18 & X2.5  & 17:31  & 17:36 & 17:36 & 12.9  & 12.3  &   \\
\tableline  2005-01-09 & M2.4  & 08:25  & 08:39 & 08:39 & 67.0  & 22.1  &   \\

\tableline
\end{tabular}
\end{center}
* The peak time is according to {\it RHESSI} light curve in the energy band of 30-50
keV.
\end{table}

\newpage

\begin{figure}
\epsscale{0.83} \plotone{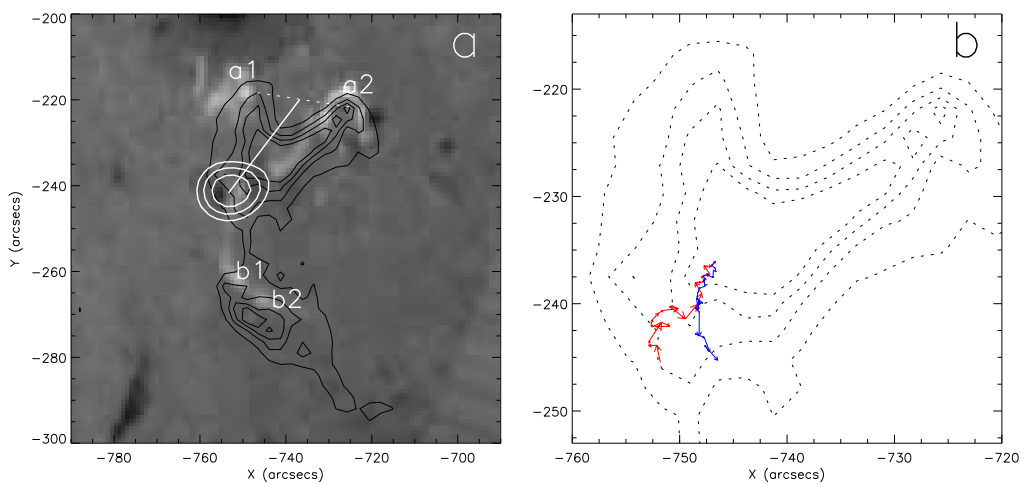} \caption{a. Multi-wavelength
presentation of the flare of 2002 September 9. Background image was
taken at Ha-1.3 A, black contours are EUV flaring loops from
SOHO/EIT observation, and white contours are from RHESSI observation
of the loop-top source in the energy band of 13-16 keV. The white
solid line shows the height of the loop-top source. b. A series of
red and blue arrows respectively shows the downward and upward
motion of the X-ray loop-top source at 13-16 keV.}
\end{figure}

\begin{figure}
\epsscale{0.83} \plotone{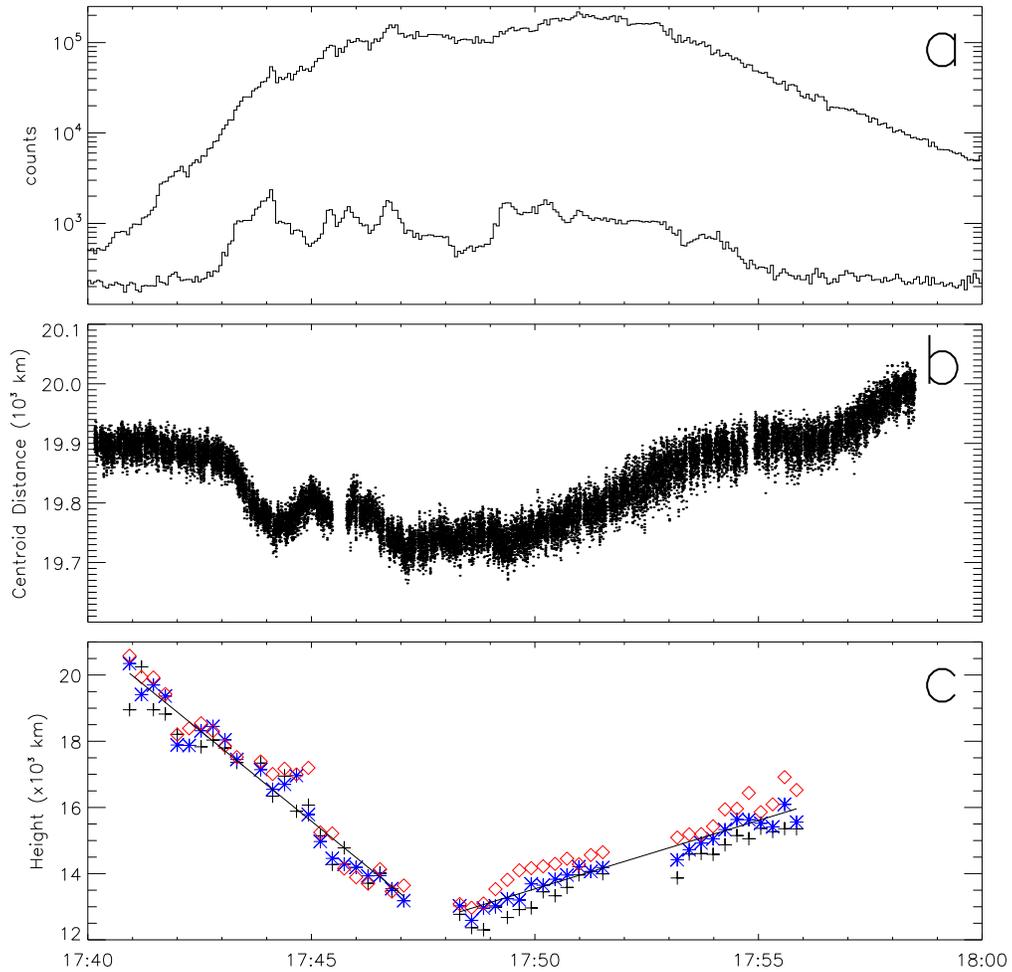} \caption{a. Light curves of HXR
emissions in energy bands of 12-25 keV and 25-50 keV. b. Time
profile for the distance between the two conjugate H$_\alpha$
kernels of the flare. c. Time profiles of the heights of the
loop-top source at three different energies (cross: 8-10 keV,
diamond: 10-13, asterisk: 13-16 keV).}
\end{figure}

\begin{figure}
\epsscale{0.92} \plotone{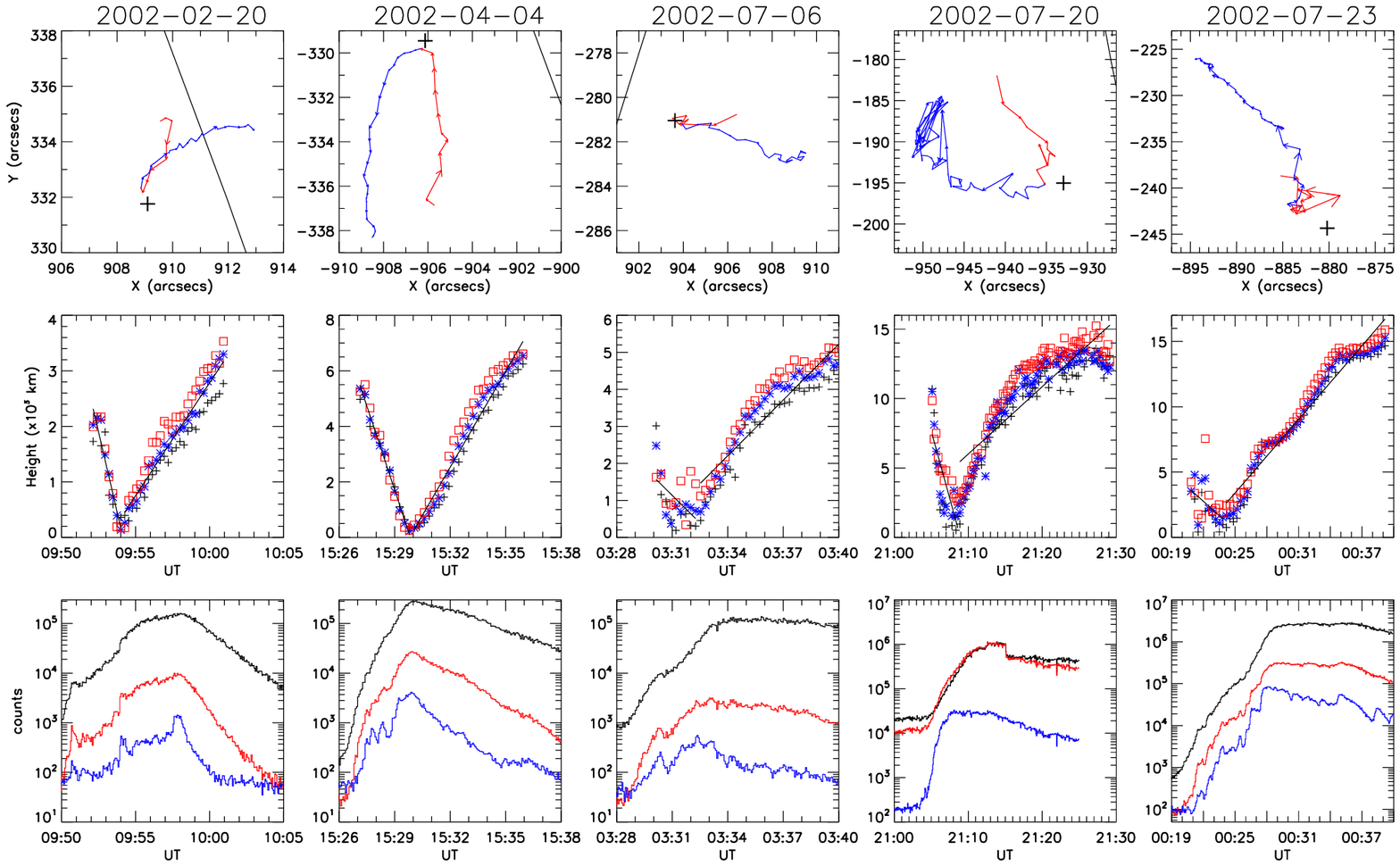} \caption{Upper panels: trajectory
of the LT source of each flare. A series of red and blue arrows
shows the downward and upward motion of each source. Middle panels:
Time profiles of the `heights' of the LT source of each flare at
different energies. (cross: 8-10 keV, diamond: 10-13, asterisk:
13-16 keV). Lower panels: RHESSI light curves in the energy range of
10-20 (black), 20-30 (red), and 30-50 keV (blue). }
\end{figure}

\begin{figure}
\epsscale{0.92} \plotone{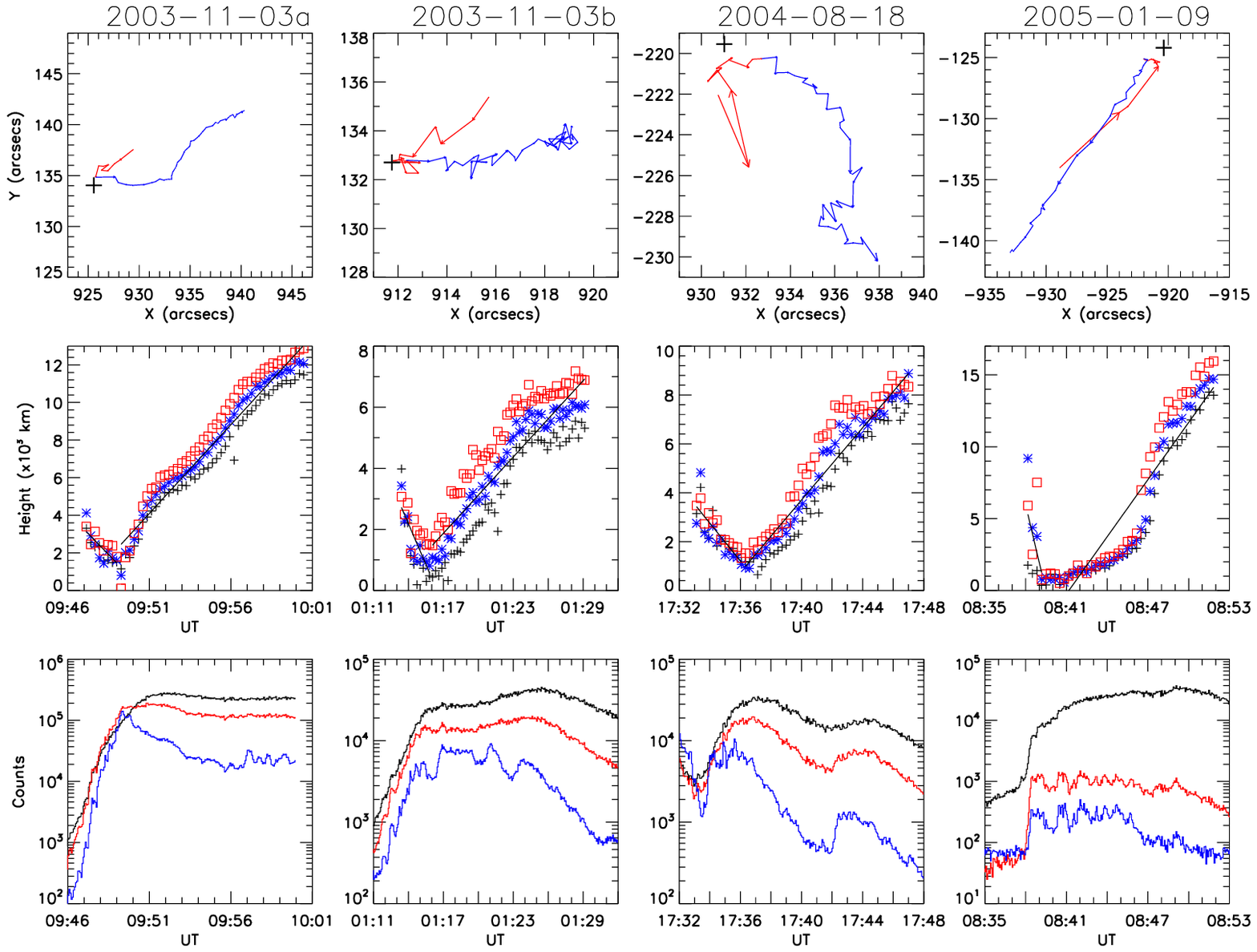} \caption{Upper panels: trajectory
of the LT source of each flare. A series of red and blue arrows
shows the downward and upward motion of each source. Middle panels:
Time profiles of the `heights' of the LT source of each flare at
different energies. (cross: 8-10 keV, diamond: 10-13, asterisk:
13-16 keV). Lower panels: RHESSI light curves in the energy range of
10-20 (black), 20-30 (red), and 30-50 keV (blue).}
\end{figure}

\end{document}